\documentclass
[aip,jmp,reprint]{revtex4-1}
\usepackage[latin9]{inputenc}
\usepackage{amsmath}
\usepackage{amssymb}
\usepackage{graphicx}
\usepackage{enumerate}

\makeatletter
\@ifundefined{textcolor}{}
{%
 \definecolor{BLACK}{gray}{0}
 \definecolor{WHITE}{gray}{1}
 \definecolor{RED}{rgb}{1,0,0}
 \definecolor{GREEN}{rgb}{0,1,0}
 \definecolor{BLUE}{rgb}{0,0,1}
 \definecolor{CYAN}{cmyk}{1,0,0,0}
 \definecolor{MAGENTA}{cmyk}{0,1,0,0}
 \definecolor{YELLOW}{cmyk}{0,0,1,0}
}

\usepackage{bm}\usepackage{amsthm}\usepackage{epstopdf}\usepackage{txfonts}

\DeclareRobustCommand{\openzero}{\leavevmode\hbox{0\kern-.55em0}}

\makeatother

\begin{document}
\title{Quantum coherence generating power, maximally abelian subalgebras, and Grassmannian Geometry}
\author{Paolo Zanardi}

\affiliation{Department of Physics and Astronomy, and Center for Quantum Information
Science \& Technology, University of Southern California, Los Angeles,
CA 90089-0484}

\author{Lorenzo Campos Venuti}

\affiliation{Department of Physics and Astronomy, and Center for Quantum Information
Science \& Technology, University of Southern California, Los Angeles,
CA 90089-0484}

%
%
%
%
%
%

\begin{abstract}
We establish a direct connection between the power of a  unitary map in $d$-dimensions ($d<\infty$) to generate quantum coherence and the geometry of the  set ${\cal M}_d$ of maximally abelian subalgebras (of the quantum system full operator algebra). This set can be  seen as a topologically non-trivial subset of the Grassmannian over linear operators.  The natural distance over the Grassmannian induces a metric structure on ${\cal M}_d$ which  quantifies the lack of commutativity between the pairs of  subalgebras.
Given a maximally abelian subalgebra one can define, on physical grounds,  an associated measure of quantum coherence. We show that the average quantum coherence generated by a unitary map acting on a uniform ensemble of quantum states in the algebra (the so-called coherence generating power of the map)  is proportional to the distance between a pair of maximally abelian subalgebras in ${\cal M}_d$ connected by the unitary transformation itself.  By embedding the Grassmannian into a projective space one can pull-back the standard Fubini-Study metric on ${\cal M}_d$  and define in this way novel geometrical measures of quantum coherence generating power. We also briefly discuss the associated differential metric structures.
\end{abstract}
\maketitle
\widetext
\section{Introduction}
The last few years have witnessed a renewed and strong interest in the quantitative theory of quantum coherence  \cite{Baumgratz-prl-2014,Levi-NJP-2014, Girolami-prl-2014,Yao-sci-rep-2016}. This has been partly motivated by the key role that this concept  plays in quantum information processing \cite{Nielsen-Book}, quantum metrology \cite{Iman-ArXiv-2016}, quantum thermodynamics \cite{Lostaglio-prx-2015,Lostaglio-nat-comm-2015}  and even in the so-called field of quantum biology \cite{Q-Biology}. A related natural question concerns with the ability of a quantum operation to generate quantum coherence. Different approaches have been explored in the literature  to quantify the coherence generating power (CGP)  of quantum operations \cite{CP1,CP2,CP3}. For a thoughtful and comprehensive review of the current efforts on the theory of quantum coherence and CGP see Ref. \cite{Adesso-review}. Also, in \cite{Dana} one can find  the most recent updates and progress on the resource theory of coherence, states and beyond.

The goal of this paper is to  develop some mathematical aspects of the  approach to CGP for finite-dimensional quantum unital operations introduced in \cite{CGP-0,CGP-1}. This is a probabilistic approach that builds on top of an analog strategy in the context of entanglement theory \cite{zanardi-praRC-2000}.
We shall  unveil the underlying geometrical and algebraic structures to the CGP measures for unitary maps defined in \cite{ CGP-0,CGP-1}.
More precisely, we will show how the formalism there  introduced can be interpreted and generalized in terms of metric structures over the space 
of maximally Abelian subalgebras (MASA) of the algebra of operators 
(the latter  being endowed with the Hilbert-Schmidt scalar product). 
The space of MASAs can be seen as a topologically non-trivial subset  of the Grassmannian manifold of $d-$dimensional
($d$ being the Hilbert space dimension) subspaces of 
of the full operator algebra and thereby inherits the Grassmannian  metric structure. 

Quite remarkably the quantitative  notion of CGP introduced in \cite{CGP-0}, on purely  physical grounds, turns out to be {\em{exactly proportional}} to the induced distance over the space of MASAs. This distance, in turn, will be shown to quantitatively measure the lack of commutativity between pairs of MASAs.   Finally by exploiting standard embeddings of the Grassmannian into projective spaces we will show how to introduce novel measures of CGP for unitaries as well as to unveil the deep geometrical origin of known ones.
 
 
 
In Sect. II we introduce the basic elements of the formalism, maximally abelian algebras and quantum coherence, and discuss their elementary properties.
In Sect. III we establish the connection between CGP measures and the geometry of the Grassmannian  over linear operators. In Sect IV we briefly analyze the associated  differential metric structure. Sect. V contains the conclusions. 

\section{Quantum coherence and maximally abelian subalgebras}
Let ${\cal H}\cong {\mathbf{C}}^d,\,(d<\infty)$ be the complex Hilbert space associated to a $d$-dimensional quantum system. The algebra of Linear operators $L({\cal H})$ is equipped with the Hilbert-Schmidt scalar product $\langle A, B\rangle:={\mathrm{tr}}( A^\dagger B)$
and $\|X\|_2:=\sqrt{\langle X, X\rangle}$.
 In the following, when $L({\cal H})$ is thought of as an Hilbert space itself with respect to this scalar product, it will be denoted by ${\cal H}_{HS}\cong {\mathbf{C}}^{d^2}$.  
 
 In the physical literature the notion of quantum coherence is usually formulated in relation to some distinguished orthonormal basis in the Hilbert space
 of the quantum system. However,  in this paper we find convenient to use a slightly more abstract, approach. We start by providing a few basic definitions and associated elementary facts.
\vskip 0.2truecm 
{\bf{Definition 0}--} A family of orthogonal projectors $B:=\{\Pi_i\}_{i=1}^m\subset L({\cal H})$ is called an orthogonal resolution of the identity (ORI)
if a) $\Pi_i  \Pi_j=\delta_{ij} \Pi_j$, b) $\sum_{j=1}^m\Pi_j=\openone$, c) $\Pi_j^\dagger = \Pi_j, \, \forall j=1,2,\ldots,m$. If all the projectors are rank one ($\Rightarrow m=d$) we will say that
$B$ is a maximal orthogonal resolution of the identity (MORI).
\vskip 0.2truecm
We would like to note  that the concept of ORI was already used in the early work on quantification of coherence \cite{aberg} in which  several of the basic questions of the resulting theory (e.g., basic state transformations, and monotones) were answered. However, in this paper the focus is on MORI's.
The set of all ORI's over ${\cal H}$ has a natural partial order ($B\le B^\prime$ iff $B^\prime$ is a {\em{refinement}} \cite{refine}
 of $B$) for which the MORI's are indeed maximal elements. 
Any orthonormal frame $(|i\rangle)_{i=1}^d$  in the Hilbert space defines a MORI i.e.~$\{\Pi_i:=|i\rangle\langle i|\}_{i=1}^d$.
A MORI on the other hand defines just an equivalence class of orthonormal frames where equivalent elements differ by permutations of the vectors
and by $U(1)$ phases. All the relevant notions and quantities  of this paper depend just on the MORI (and not on the specific frame in the equivalence class).  

 Let us next consider  one of the main objects of this paper: the algebra of operators which are diagonal in the representation associated to any frame in the equivalence class of $B$.
\vskip 0.2truecm 
{\bf{Definition 1}--} Given an MORI $B=\{\Pi_i\}_{i=1}^d$ we define the associated $d$-dimensional abelian subalgebra (ASA) of $L({\cal H})$ by:
${\cal A}_B:=\lbrace\sum_{j=1}^d \lambda_j \Pi_j\,/\, (\lambda_j)_{j=1}^d\in{\mathbf{C}}^d\rbrace\subset {\cal H}_{HS}$.
The  map 
\begin{equation}
{\cal D}_B\colon {\cal H}_{HS}\rightarrow {\cal H}_{HS}/X\mapsto \sum_{j=1}^m \Pi_j X\Pi_j
\label{D_B}
\end{equation}
 is an orthogonal projection (in ${\cal H}_{HS}$) whose range is  ${\cal A}_B$. 
\vskip 0.2truecm
 Clearly ${\cal A}_B$ is closed under hermitean conjugation.
At the physical level the algebra projection  ${\cal D}_B$ is the measurement map associated to the MORI $B$ and it is a completely positive (CP) trace-preserving unital map i,e., $ {\cal D}_B(\openone)=\openone$.
Crucially, the spaces ${\cal A}_B$  are {\em{maximal}} ASAs (MASA)  in the sense that they are  not a proper subalgebras of any other abelian one.
This basically follows from the fact that the  map $B\mapsto {\cal A}_B$ between ORIs and subalgebras of $L({\cal H})$  is a morphism of partially ordered sets i.e., 
 $B\le B^\prime\Rightarrow {\cal A}_B\subset  {\cal A}_{B^\prime}$.
%
\vskip 0.2truecm
{\bf{Proposition 1--}}  
 Let ${\cal M}_d$ denote the family of MASAs over ${\cal H}_{HS}$.

{\bf{i)}} The correspondence $B\mapsto{\cal A}_B$ is a bijection between  the set of all MORIs and ${\cal M}_d$.

{\bf{ii)}} The action $U(d)\times{\cal M}_d\rightarrow {\cal M}_d\colon (U,{\cal A})\mapsto {\cal U}({\cal A}):=\{{\cal U}(X):=UXU^\dagger\,/\,X\in {\cal A}\}$
is transitive. Moreover, 
\begin{equation}
{\cal M}_d\cong  \frac{X_d}{{\cal S}_d},\quad X_d=\frac{U(d)}{U(1)^d}
\label{M_d}
\end{equation}
where ${\cal S}_d$ denotes the permutation group of $d$-objects.

{\em{Proof.--}} 
{\bf{i)}} 
We have to show that a)  If $B$ is a MORI then ${\cal A}_B\in {\cal M}_d$ and b)  if ${\cal A}\in {\cal M}_d$ 
then there exists a MORI $B$ such that ${\cal A}={\cal A}_B$. Moreover the correspondence $B\mapsto {\cal A}_B$ is one-to-one.
 
Let $B=\{\Pi_i=|i\rangle\langle i|\}_{i=1}^d$ and suppose ${\cal A}_B\subset {\cal A}$ where $\cal A$ is an abelian subalgebra of $L({\cal H})$. 
If $X\in{\cal A}$ then $[X,A]=0\,\forall A\in{\cal A}_B$.
In particular $A\Pi_i=\Pi_i A\,\forall i$ whence   $A|i\rangle =\langle i|A|i\rangle\,|i\rangle\,\forall i$.
This shows that $A$ is $B$-diagonal i.e., $A\in{\cal A}_B\,\forall A\in{\cal A}$ and therefore ${\cal A}={\cal A}_B$.
{{b)}} Suppose ${\cal A}\subset L({\cal H})$ is a MASAs. Any $A \in{\cal A}$ can be written as sum of an hermitean
and an anti-hermitean commuting parts. Moreover since all $A$'s commute there exists an ORI $\{Q_j\}_j$
such that $A=\sum_j \alpha_j  Q_j$ (joint diagonal form for all elements of ${\cal A}$). Now, all the $Q_j$'s have to be
one-dimensional i.e., $\{Q_j\}_j$ has to be a MORI. In fact, if it were not so it would exist at least one $j_0$ such that $Q_{j_0}$
is higher-dimensional and therefore there it would exist a $S\in L({\cal H})$ which is {\em{non}}-diagonal but still commutes with all $Q_j$'s
(and therefore with all elements in $\cal A$). For example one may consider
a unitary map $S$ which acts like the identity everywhere but on the range of $Q_{j_0}$ where it is a non-trivial unitary.
The algebra generated by $\cal A$ and $S$ is still abelian and strictly contains $\cal A$. This shows that, unless all the $Q_j$ are rank one
the algebra $A$ is not a MASA. In conclusion if $\cal A$ is a MASA then it is generated by a MORI i.e., the map $B\mapsto {\cal A}_B$ is surjective.

Finally let us assume ${\cal A}_{\tilde{B}}={\cal A}_B$. This implies in particular that $\tilde{\Pi}_i=\sum_{j=1}^d p_j\Pi_j\,(\forall i)$.
The spectrum of the LHS of this identity is $\{0,1\}$ while the one of the RHS is $\{p_j\}_{j=1}^d$. 
Therefore $p_j\in\{0,1\}$ because they form a probability distribution, $\forall i$  there  exists a $j=j(i)\in\{1,\ldots,d\}$ such that $\tilde{\Pi}_i= \Pi_{j(i)}$. This shows that the elements of $\tilde{B}$ are just a permutation of those of $B$ i.e., $B=\tilde{B}$. 
This amounts to prove that $B\mapsto {\cal A}_B$ is injective.

{\bf{ii)}} Since ${\cal M}_d$ is, as a set, the same as the set of all MORIs we will focus on the structure of the latter.
Let us consider the set $X_d=\{(\Pi_i)_{i=1}^d\}$ of all {\em{ordered}} $d$-tuples of projectors $\Pi_i$'s such that
$\{\Pi_i\}_{i=1}^d$ is a MORI. By defining the ${\cal S}_d$-action on $X_d$ by $(\sigma, (\Pi_i)_{i=1}^d)\mapsto (\Pi_{\sigma(i)})_{i=1}^d$
then clearly the set of MORIs is nothing but $X_d/{\cal S}_d$. Now, if $x:=\{(\Pi_i=|i\rangle\langle i|)_{i=1}^d\}, 
\tilde{x}=\{(\tilde{\Pi}_i=|\tilde{i}\rangle\langle \tilde{i}| )_{i=1}^d\}\in X_d$ then $U=\sum_{i=1}^d |\tilde{i}\rangle\langle i|=(U^\dagger)^{-1}$ maps 
one into the other by $x\mapsto U\cdot x:=(U\Pi_i U^\dagger)_{i=1}^d=\tilde{x}$. This means that $U(d)$ acts transitively over $X_d$
as well as on ${\cal M}_d$ (forgetting the order).
On the other hand the stabilizer of $x$ is given by $\{\sum_{i=1}^d \chi_i\Pi_i\,/\, \chi_i\in U(1), \,i=1,\dots, d\}\cong U(1)^d$.
Then $X_d\cong U(d)/U(1)^d$  follows from the standard identification of the $U(d)$-homogeneous space $X_d$
with the coset space obtained by dividing the group $U(d)$ by the stabilizer subgroup. This concludes the proof.
$\hfill\Box$

\vskip 0.2truecm
The space $X_d$ in Eq.~(\ref{M_d}) can be seen as the compact, simply-connected manifold of orthogonal  {\em{full-flags}} \cite{flag}.
This implies  that ${\cal M}_d$ is topologically non-trivial as its fundamental group is  isomorphic to  ${\cal S}_d$.
Indeed $\pi_1({\cal M}_d)=\pi_1( \frac{X_d}{{\cal S}_d})\cong {\cal S}_d$.


Having introduced the basic algebraic and geometrical objects of our formalism we now turn to physical concepts \cite{CGP-0}.
\vskip 0.2truecm 
{\bf{Definition 2}--}  Given  the MORI  $B=\{\Pi_i\}_{i=1}^d$  we define:%
\begin{enumerate}[a)]
\item The {\em{$B$-incoherent states}} as the set of quantum states in ${\cal A}_B$ i.e.,
$I_B:=\{\sum_{j=1}^d p_j\Pi_j\,/\, p_j\ge 0,\,\sum_{j=1}^d p_j=1\}\subset {\cal A}_B\in{\cal M}_d.
$

\item Given the quantum state $\rho$ we define its $B$-coherence by 
\begin{equation}
c_B(\rho):=\inf_{X\in{\cal A}_B}\|\rho-X\|_2^2=\|\rho-{\cal D}_B(\rho)\|_2^2=\|{\cal Q}_B(\rho)\|_2^2.
\label{c_B}
\end{equation}
\item A unital $CP$-map ${\cal T}\colon {\cal H}_{HS}\rightarrow{\cal H}_{HS}$ is called {\em{incoherent}} iff $[{\cal T},{\cal D}_B]=0$.
\end{enumerate}
\vskip 0.2truecm
A couple of comments are here in order. 1) The definition above relies on the Hilbert-Schmidt norm $\|\bullet\|_2$ this, on the one hand, leads to a somewhat simplified theory of quantum coherence as naturally restricts the set of allowed  operations to unital ones (for which the Hilbert-Schmidt norm is not increasing). On the other hand 
the simpler properties of $\|\bullet\|_2$ allows one to obtain a wealth of rigorous analytical that can be hardly obtained by more information-theoretic motivated choices e.g., the trace norm $\|\bullet\|_1.$  2) We also note that our definition of incoherent operations above falls in the class of dephasing-covariant incoherent operation as per the categorification of Ref. \cite{Adesso-review} (see Table II therein).

The set $I_B$ is clearly a $(d-1)$-dimensional simplex.
The first equality in Eq.~(\ref{c_B}) stems from the fact that ${\cal A}_B$ is a (closed) linear subspace of ${\cal H}_{HS}$ and that ${\cal D}_B$ is the corresponding orthogonal projection on it. This equality also shows that $c_B(\rho)=\inf_{\sigma\in I_B}\|\rho-\sigma\|_2^2.$ 
The second equality simply defines the complementary projection ${\cal Q}_{B} := \mathbf{1} - {\cal D}_B$. 
Notice that, from c) above, an incoherent map ${\cal T}$ is such that ${\cal T}({\cal A}_B)\subset {\cal A}_B$. The latter condition, which can be written
as ${\cal D}_B{\cal T}{\cal D}_B={\cal T}{\cal D}_B,$ is a weaker notion of incoherence coinciding with c) for  normal maps $\cal T$ \cite{CGP-0}.

Next we show that  Eq.~(\ref{c_B}) defines a good coherence measure for unital maps and that it can also be seen as quantitative measure of the lack of commutativity between the state $\rho$ and ${\cal A}_B\in{\cal M}_d$.

\vskip 0.2truecm 
{\bf{Proposition 2--}} 
{\bf{i)}} The map $\rho\mapsto c_B(\rho)$ over quantum states $\rho$ defined by Eq.~(\ref{c_B}) defines a good {\em{coherence measure}} i.e.,
$c_B(\rho)=0$ if $\rho\in I_B$ and  $c_B({\cal T}(\rho))\le c_B(\rho)$ for ${\cal T}$ incoherent.

{\bf{ii)}} Let $B=\{\Pi_i\}_{i=1}^d$ be a MORI and $\rho$ a quantum state, then one has
$c_B(\rho)=\frac{1}{2}\sum_{j=1}^d\|[\Pi_j, \rho]\|_2^2$.

{\em{Proof.--}} 
{\bf{i)}} This was proved in \cite{CGP-0} we report the proof here for completeness.
First, from (\ref{c_B}) and the definition of $I_B$ one has $c_B(\rho)=0\Leftrightarrow {\cal Q}_B(\rho)=0 \Leftrightarrow \rho={\cal D}_B(\rho)\Leftrightarrow\rho\in I_B$. Second, $c_B({\cal T}(\rho))=\|{\cal Q}_B{\cal T}(\rho)\|_2^2=\|{\cal T}{\cal Q}_B(\rho)\|_2^2\le 
\|{\cal Q}_B(\rho)\|_2^2=c_B(\rho)$. Here we have used that for incoherent maps $[{\cal T}, {\cal Q}_B]=0$ and that the Hilbert-Schmidt
is monotonic under unital maps ${\cal T}$ i.e., $\| {\cal T}(X)\|_2\le \|X\|_2,\,(\forall X\in {\cal H}_{HS})$.

{\bf{ii)}}
A simple computation shows:
$$\|[\Pi_j,\rho]\|_2^2  =\|\Pi_j\rho-\rho\Pi_j\|_2^2 
={\mathrm{tr}}\left( \Pi_j\rho^2\Pi_j +\rho\Pi_j\rho- \Pi_j\rho\Pi_j\rho 
-\rho\Pi_j\rho\Pi_j \right)  
=2\, \left[  {\mathrm{tr}}(\rho^2\Pi_j)-  {\mathrm{tr}}(\Pi_j\rho\Pi_j\rho ) \right].
$$

Summing over $i$ one obtains
$\sum_{i=1}^d \|[\Pi_j,\rho]\|_2^2=2\,(\|\rho\|_2^2-\|{\cal D}_B(\rho)\|_2^2=2\,\|{\cal Q}_B(\rho)\|_2^2=2\,c_B(\rho)$.
Here we have used the definition of ${\cal D}_B$ in Eq.~(\ref{D_B})
 and the fact that $\langle {\cal D}_B(\rho), \rho\rangle=\langle {\cal D}_B(\rho), {\cal D}_B(\rho)\rangle =\|{\cal D}_B(\rho)\|_2^2$.
$\hfill\Box$
\vskip 0.2truecm
 This result shows that the geometric notion of distance, the algebraic one of non-commutativity and the physical one of quantum coherence are tightly  tied together at the level of a single quantum state $\rho$. In the following we will demonstrate that this connection holds at the level of the coherence generating power of unitary maps  and pairs of MASAs.

\section{Coherence power and Grassmannian geometry}
Once $B$-coherence is defined one can ask the question about the ability of a unital CP-map to generate it. Here we will follow the probabilistic approach
advocated in  Ref. \cite{CGP-0}. The idea is that the coherence generating power (CGP) of a map $\cal T$ is the average coherence --as quantified by (\ref{C_B}) -- generated by $\cal T$ acting on a uniform  ensemble of incoherent states. More precisely, let us now consider the uniform probability measure over $I_B$ \cite{CGP-0} and denote by ${\mathbf{E}}_{\mathrm{unif}: I_B}\left[\bullet\right]$
the corresponding expectation.
\vskip 0.2truecm
{\bf{Definition 3}--} Given the unital CP-map ${\cal T}\colon {\cal H}_{HS}\rightarrow {\cal H}_{HS}$ we define its $B$-coherence generating power 
(CGP) by
\begin{equation}
C_B({\cal T}):={\mathbf{E}}_{\mathrm{unif}: I_B}\left[ c_B({\cal T}(\rho))\right] 
\label{C_B}
\end{equation}
\vskip 0.2truecm
This approach to CGP is based on probabilistic averages as opposed to optimizations over set of states and/or protocols see e.g., \cite{Dana}.
Clearly, this choice makes harder to envisage a direct operational and information-theoretic meaning of (\ref{C_B}). However, this strategy, along with the nice algebraic properties of the Hilbert-Schmidt norm, allows one to  find explicit analytical results for {\em{ arbitrary (unital) maps and dimensions.}} In fact, in Ref. \cite{CGP-0} we have proven the following fact
\vskip 0.2truecm 
{\bf{Proposition 3--}} Let $B=\{\Pi_i:=|i\rangle\langle i|\}_{i=1}^d$ be a MORI  and $\cal T$ a unital CP over ${\cal H}_{HS}$ then 
$C_B({\cal T})=N_d \sum_{j=1}^d\| {\cal Q}_B {\cal T}(\Pi_j)  \|_2^2,\, N_d^{-1}:= d(d+1)$.
In particular for unitary CP-maps ${\cal U}(X):= UX U^\dagger \,(U\in U(d))$ one has
\begin{equation}
C_B({\cal U})=N_d (d-  \sum_{i,j=1}^d |\langle i|U|j\rangle|^4)
\label{U-C_B-CGP-0}
\end{equation}

{\em{Proof.--}} See Prop. 4 in Ref. \cite{CGP-0} $\hfil\Box$
\vskip 0.2truecm

It is now  important to observe that ${\cal M}_d$ is a subset  of the {\em{Grassmannian}} of $d$-dimensional linear subspaces $W$'s
\begin{equation}
{\cal G}_d({\cal H}_{HS}):=\{W\subset {\cal H}_{HS}\,/ \mathrm{dim} W=d\}\supset {\cal M}_d
\label{grass}
\end{equation}
This is a differentiable manifold with (real) dimension $d^2 (d^2-2)$. 
Now we would like to show that the CGP (\ref{U-C_B-CGP-0}) has an underlying origin at the level of the geometry of ${\cal M}_d$. 
The first step is to observe that, since MASAs belong to the Grassmannian, the set ${\cal M}_d$ inherits the metric structure of the latter.  
\vskip 0.2truecm
{\bf{Definition 4--}}
Let  ${\cal A}_B, {\cal A}_{\tilde{B}}\in{\cal M}_d$  we define a metric structure over ${\cal M}_d$ by
\begin{equation}
D({\cal A}_B, {\cal A}_{\tilde{B}}):=\|{\cal D}_B-{\cal D}_{\tilde{B}}\|_{HS}.
\label{dist}
\end{equation}
where $\|\bullet\|_{HS}$ denotes the Hilbert-Schmidt norm over $L({\cal H}_{HS})$.
\vskip 0.2truecm
It  is  a well-known fact that the distance between subspaces in a Grasmannian   can be taken to be the (Hilbert-Schmidt) distance between the corresponding orthogonal projections; Eq.~(\ref{dist}) is just the particular case for elements of ${\cal M}_d\subset {\cal G}_d({\cal H}_{HS})$.
Notice that the distance (\ref{dist}) is  invariant under the $U(d)$ action over ${\cal M}_d$ i.e., $ D({\cal U}({\cal A}_B), {\cal U}({\cal A}_{\tilde{B}}))=
D({\cal A}_B, {\cal A}_{\tilde{B}})$ \cite{uni-inv}.

One can now  establish a direct connection between the distance (\ref{dist}) 
between MASAs and the, apparently totally unrelated,  CGP of unitaries (\ref{U-C_B-CGP-0}). 
The following  proposition  contains  some of the key conceptual as well as technical results of this paper
\vskip 0.2truecm
{\bf{Proposition 4--}} 
{\bf{i)}} Let ${\cal U}$ be a unitary CP-map, then
\begin{equation}
C_B({\cal U})=\frac{N_d}{2}\, D^2({\cal A}_B, {\cal U}({\cal A}_{{B}}))
\label{C_B-dist}
\end{equation}

{\bf{ii)}}
$D^2({\cal A}_B, {\cal A}_{\tilde{B}})=\sum_{i,j=1}^d\|[\Pi_i, \tilde{\Pi}_j]\|_2^2$

{\em{Proof.--}} 
{\bf{i)}} We first show how to compute traces over $L({\cal H}_{HS})$ for maps of the form ${\cal F}\colon X\mapsto A X B$ where $A,B\in  L({\cal H})$.
By definition $\mathrm{Tr}\,({\cal F})=\sum_{s=1}^{d^2} \langle X_s, {\cal F}(X_s)\rangle$ where $\{X_s\}_{s=1}^{d^2}$ is an orthonormal basis in ${\cal H}_{HS}$.
If $\{|i\rangle\}_{i=1}^d$ is an orthonormal basis of $\cal H$ let us pick $X_s=|l\rangle\langle m|,\,s=(l,m),\,(l,m=1,\ldots, d)$. Whence,
${\mathrm{Tr}}\,({\cal F})=\sum_{l,m=1}^d {\mathrm{tr}}( |m\rangle\langle l|A |l\rangle\langle m|B)=(\sum_{l=1}^d \langle l|A|l\rangle)
(\sum_{m=1}^d \langle m|B|m\rangle)={\mathrm{tr}}(A){\mathrm{tr}}(B)$. 
Let ${\cal A}_B, {\cal A}_{\tilde{B}}\in{\cal M}_d$ associated to the MORIs $B=\{\Pi_i=|i\rangle\langle i|\}_{i=1}^d$ and $\tilde{B}=\{\tilde{\Pi}_i=|\tilde{i}\rangle\langle \tilde{i}|\}_{i=1}^d$ 
respectively. Now $D^2({\cal A}_B, {\cal A}_{\tilde{B}})=\|{\cal D}_B-{\cal D}_{\tilde{B}}\|_{HS}^2=\|{\cal D}_B\|_{HS}^2+\|{\cal D}_{\tilde{B}}\|_{HS}^2-2{\mathrm{Tr}}({\cal D}_B{\cal D}_{\tilde{B}})$.
The first term can be written as  ${\mathrm{Tr}}({\cal D}_B^2)={\mathrm{Tr}}(\sum_{i,j=1}^d \Pi_i\Pi_j\bullet\Pi_j\Pi_j)=\sum_{i,j=1}^d  {\mathrm{Tr}}(\Pi_i\Pi_j)^2=\sum_{i,j=1}^d \delta_{ij}  {\mathrm{Tr}}(\Pi_j)^2=d$. The same is true for the second term. Let us now turn to the last term
${\mathrm{Tr}}({\cal D}_B{\cal D}_{\tilde{B}})=\sum_{i,j=1}^d {\mathrm{Tr}}(\Pi_i\tilde{\Pi}_j\bullet\tilde{\Pi}_j\Pi_i)=\sum_{i,j=1}^d 
{\mathrm{tr}}(\Pi_i\tilde{\Pi}_j)^2=\sum_{i,j=1}^d{\mathrm{tr}}(|i\rangle\langle i|\tilde{j}\rangle\langle\tilde{j}|)^2=\sum_{i,j=1}^d |\langle i|\tilde{j}\rangle|^4$.
Adding the three terms  one gets
\begin{equation}
D^2({\cal A}_B, {\cal A}_{\tilde{B}})=2(d-\sum_{i,j=1}^d |\langle i|\tilde{j}\rangle|^4).
\label{ii}
\end{equation}
Now  set $|\tilde{j}\rangle:= U|j\rangle$ in the last equation  and compare with Eq.~(\ref{U-C_B-CGP-0}). 

{\bf{ii)}} It is a direct computation. $\sum_{i,j=1}^d\|[\Pi_i, \tilde{\Pi}_j\|_2^2=\sum_{i,j=1}^d\|\Pi_i \tilde{\Pi}_j-\tilde{\Pi}_j\Pi_i\|_2^2=
\sum_{i,j=1}^d {\mathrm{tr}}(\Pi_i \tilde{\Pi}_j\Pi_i+\tilde{\Pi}_j\Pi_i\tilde{\Pi}_j -\Pi_i\tilde{\Pi}_j\Pi_i\tilde{\Pi}_j- \tilde{\Pi}_j\Pi_i\tilde{\Pi}_j\Pi_i)=
2\sum_{i,j=1}^d \left(  {\mathrm{tr}}(\Pi_i \tilde{\Pi}_j) -  {\mathrm{tr}}(\Pi_i \tilde{\Pi}_j \Pi_i\tilde{\Pi}_j) \right)=2(d-\sum_{i,j=1}^d |\langle i|\tilde{j}\rangle|^4)$. Comparing with Eq.~(\ref{ii}) concludes the proof. 

An alternative proof can be obtained by setting in {\bf{Prop. 2}}
$\rho=\sum_{k=1}^d p_k {\cal U}(\Pi_k)$, expanding the commutators norms and using ${\mathbf{E}}_{{\mathrm{unif:}} I_B}[p_i p_k]=N_d(1+\delta_{ik})$ \cite{CGP-0}.
 $\hfill\Box$
\vskip 0.2truecm

Eq.~(\ref{C_B-dist}) allows one to immediately and elegantly derive several properties of the CGP of unitaries (\ref{U-C_B-CGP-0}).
First, the only unitaries with zero CGP are those which fix ${\cal A}_B$ i.e., the incoherent ones [see {\bf{Def. 2}} c)]. 
Second, if ${\cal W}$ is $B$-incoherent because of the unitary invariance of the distance (\ref{dist}) one has that $D({\cal A}_B, {\cal W} {\cal U}({\cal A}_{{B}}))= D({\cal W}({\cal A}_B), {\cal W} {\cal U}({\cal A}_{{B}}))= D({\cal A}_B, {\cal U}({\cal A}_{{B}}))$. Now Eq.~(\ref{C_B-dist}), implies $C_B({\cal U})=C_B({\cal W}{\cal U})$. Namely, the CGP of a map is invariant under postprocessing by incoherent unitaries \cite{CGP-0}.
Invariance under pre-processing by incoherent maps is trivial from (\ref{C_B-dist}). Third, from
$D({\cal A}_B, {\cal U}({\cal A}_{{B}}))=D({\cal U}^\dagger({\cal A}_B), {\cal U}^\dagger {\cal U}({\cal A}_{{B}}))=
D({\cal U}^\dagger({\cal A}_B),{\cal A}_{{B}})=D({\cal A}_{{B}}, {\cal U}^\dagger({\cal A}_B))$ and  (\ref{C_B-dist}) one gets
$C_B(U)=C_B(U^\dagger).$ The CGP of a unitary is equal to the one of its inverse.


At the conceptual level these results  demonstrate  that the physical concept of CGP, the metric structure of the Grassmannian ${\cal G}_d({\cal H}_{HS})$ (more precisely of ${\cal M}_d$) and  quantum non-commutativity are profoundly connected to each other.
In words: the $B$-coherence generating power of a unitary map $\cal U$ is proportional to the Grassmannian distance between the input $B$-diagonal algebra ${\cal A}_B$ and its image under $\cal U$. This distance, in turn, can be quantitatively identified with the lack of commutativity between these two algebras \cite{commA-B}.  It is important to stress that, in the light of the results of \cite{CGP-0}, the latter geometrical and algebraic properties can be directly measured by a quantum experiment i.e., Grassmannian metric and non-commutativity  are endowed with a physical as well as  operational meaning.

In passing we mention that relation (\ref{C_B-dist}) suggests a straightforward path to extend the notion of CGP to infinite dimensions. Indeed one can replace the Hilbert-Schmidt norm in Eq.~(\ref{dist}) by any unitary invariant norm for CP-maps  and then {\em{define}} the CGP of a unitary map as the corresponding distance between ${\cal A}_B$ and ${\cal U}({\cal A}_B).$ However, for $d=\infty$
the characterization of the set of MASAs is a much more challenging task and it lies beyond the scope of this paper.

Another appealing feature of the framework  here discussed is that it also allows one to introduce novel measures for CGP of unitaries with an underlying geometrical meaning. To this aim it is useful to introduce one more definition.
\vskip 0.2truecm
{\bf{Definition 5--}} Given a pair of {\em{ordered}}  MORIs $B_<=:(\Pi_i=|i\rangle\langle i|)_{i=1}^d, B_<=:(\tilde{\Pi}_i=|\tilde{i}\rangle\langle \tilde{i}|)_{i=1}^d\in X_d$ we define the  associated $d\times d$ non-negative {\em{overlap}} matrix by
$\hat{O}_{i,j}(B_<,\tilde{B}_<):=\langle \Pi_i,\tilde{\Pi}_j\rangle=|\langle i|\tilde{j}\rangle|^2\ge 0,\;(i,j=1,\ldots,d)$.
In particular if $B_<$ and $\tilde{B}_<$ are connected by the unitary $U$ i.e., $\tilde{B}_<=BU_<:=({\cal U}(\Pi_i))_{i=1}^d$ 
 we define $\hat{X}_{B_<}(U):=\hat{O}(B_<, BU_<)$ [$\hat{X}_{B_<}(U)_{i,j}=|\langle i|U|j\rangle|^2,\,(i,j=1,\ldots,d)$]
\vskip 0.2truecm
We first notice that $\hat{O}$ is doubly-stochastic for any pair $(B_<,\tilde{B}_<).$ Indeed, 
summing over $j$ one finds $\sum_{j=1}^d \hat{O}_{i,j}(B_<,\tilde{B}_<)=\sum_{j=1}^d \langle \Pi_i,\tilde{\Pi}_j\rangle
=\langle \Pi_i,\openone\rangle={\mathrm{tr}}\,\Pi_i=1,(\forall i)$. The same result is obtained by summing over $i.$ 
 Lets us remind the reader that, from the Proof of {\bf{ii)}} of {\bf{Prop 1}}, the set $X_d$ of unordered MORIs
is acted upon by ${\cal S}_d$ via $(\Pi_i)_{i=1}^d\times \sigma\mapsto (\Pi_{\sigma(i)})_{i=1}^d$ and that a MORI in ${\cal M}_d$ is just an ${\cal S}_d$-equivalence class $[(\Pi_i)_{i=1}^d] $[see Eq.~(\ref{M_d})].
The next proposition shows the other properties of the overlap matrix and how it can be used to define novel metric structures over ${\cal M}_d$ as well as CGP measures for unitaries. 
\vskip 0.2truecm
{\bf{Proposition 5.--}} 
 {\bf{i)}} The real-valued functions over $X_d\times X_d$ defined by $\|\hat{O}(B_<,\tilde{B}_<)\|_2$ and $|\det\, \hat{O}(B_<,\tilde{B}_<)|$
 depend only on the  ${\cal S}_d$-equivalence classes  $B=[B_<]$ and  $\tilde{B}=[\tilde{B}_<]$   i.e., they are functions
 over ${\cal M}_d\times{\cal M}_d.$ Moreover
\begin{equation}
1\le \|\hat{O}(B_<,\tilde{B}_<)\|_2^2={\mathrm{Tr}} ({\cal D}_B {\cal D}_{\tilde{B}})\le d,\quad
|\det \hat{O}(B_<,\tilde{B}_<)|=1\Leftrightarrow
B=\tilde{B}.
\label{overlap-prop}
\end{equation}
{\bf{ii)}} The function $D_{FS}\colon {\cal M}_d\times{\cal M}_d\rightarrow {\mathbf{R}}^+$ given by
\begin{equation}
 D_{FS}({\cal A}_B, {\cal A}_{\tilde{B}})):=\cos^{-1}(|\det \hat{O}(B_<,\tilde{B}_<)|), 
 \label{D_FS}
 \end{equation}
where $B_<$ and $\tilde{B}_<$ are any ordered MORIs in the ${\cal S}_d$-equivalence classes $B=[B_<]$ and  $\tilde{B}=[\tilde{B}_<]$
 respectively, defines a unitary invariant metric over ${\cal M}_d$.

{\bf{iii)}}  The following  functions define  good CGP measures.
\begin{align}
\tilde{C}_{B}(U) &:=D_{FS}({\cal A}_B, {\cal U}({\cal A}_{B}))=\cos^{-1}(|\det \hat{X}_{B_<}(U)|) \label{Cs} \\
 \varphi_B(U) &:=-\frac{1}{d}\ln |\det \hat{X}_{B_<}(U)|
\label{additive}
\end{align}
where $B_<$ is  any ordered MORI in the ${\cal S}_d$-equivalence class  $B=[B_<].$

{\em{Proof.--}} {\bf{i)}}  
It can easily  checked that if one reorders the elements in $B_<$ and $\tilde{B}_<$ the overlap matrix transforms according to $\hat{O}\mapsto Q\hat{O} P^T$ where $P$ and $Q$ are unitary permutation  matrices. 
From which one immediately obtains the first part of {\bf{i)}}.
%
The first equality in Eq.~(\ref{overlap-prop}) reads $\sum_{i,j=1}^d |\langle i |\tilde{j}\rangle |^4={\mathrm{Tr}} ({\cal D}_B {\cal D}_{\tilde{B}})$ which has been already proven in the proof of {\bf{Prop. 4}} (see lines above Eq.~(\ref{ii}). The range indicated follows from the fact
that this norm is the sum of the purities of $d$ probability vectors in $d$-dimensions (see below).
Let us now turn to the second equality in Eq.~(\ref{overlap-prop}).
Let $\hat{O}=W O_D V^\dagger$ be a Singular Value Decomposition of $\hat{O}$ with $W$ and $V$ unitaries and $O_D=\mathrm{diag}\,(\lambda_1,\ldots,\lambda_d)$
the diagonal matrix of the singular values of $\hat{O}$. One has that $|\det(\hat{O})|=\det({O}_D)=\prod_{i=1}^d\lambda_i$.
The squares of the $\lambda_i$'s on the other hand are the eigenvalues of the doubly-stochastic matrix $\hat{O}^T \hat{O}$
whence $0\le \lambda_i\le 1,\,(\forall i)$. It follows that $|\det(\hat{O})|=1$ iff $\lambda_i=1\,(\forall i)$.
This in turn is equivalent to $\| \hat{O}\|_2^2={\mathrm{tr}}(\hat{O}^T \hat{O})=\sum_{i,j=1}^d \hat{O}_{ij}^2=\sum_{i=1}^d\lambda_i^2=d$.
Since $\hat{O}_i:=(\hat{O}_{ij})_{j=1}^d$ (from double-stochasticity) are probability vectors $\forall i$ the former equality is possible
iff $\sum_{j=1}^d \hat{O}_{ij}^2=1\,(\forall i)$ i.e., all the $\hat{O}_i$'s are {\em{pure}}. 
This means that $\forall i\exists j=\sigma(i)$ such that $\hat{O}_{i,j}=\delta_{j,\sigma(i)}$. Moreover since $\sum_{j=1}^d \hat{O}_{j,\sigma(i)}=1$
one sees that $\sigma$ must be  in ${\cal S}_d$. In summary $|\det(\hat{O})|=1$ iff $\exists \sigma\in{\cal S}_d$ such that
$\hat{O}_{ij}=\delta_{j,\sigma(i)}$ this amounts to say that ${\Pi}_i=\tilde{\Pi}_{\sigma(i)},$ for some permutation $\sigma$ and $\forall i$
i.e., $B=\tilde{B}$.

{\bf{ii)}} In order to show that (\ref{D_FS}) defines a distance function over ${\cal M}_d$
we resort to the well-known Pl\"ucker embedding \cite{plucker}.
For  MASAs ${\cal A}_B$ with $B=\{\Pi_i=|i\rangle\langle i|\}_{i=1}^d$ this embedding takes the form
\begin{equation}
\psi\colon {\cal M}_d\rightarrow {\mathbf{P}}(\bigwedge_{i=1}^d {\cal H}_{HS})\,/\,{\cal A}_B\mapsto [\wedge_{i=}^d \Pi_i],
\label{Plucker}
\end{equation}
where $\wedge_{i=1}^d \Pi_i
:=\frac{1}{d!}\sum_{\sigma\in{\cal S}_d}(-1)^{|\sigma|}\otimes_{k=1}^d \Pi_{\sigma(k)},$
 and $[\bullet]$ denotes the projective equivalence class.
The standard Fubini-Study metric of the projective space $d_{FS}([v],[w]):=\cos^{-1}(|\langle v| w\rangle|)$ is given by
$d_{FS}(\psi({\cal A}_B),\psi({\cal A}_{\tilde{B}}))$. Using (\ref{Plucker}) and the standard properties
of anti-symmetrized tensor products one  finds  
\begin{equation}
|\langle \psi({\cal A}_B), \psi({\cal A}_{\tilde{B}})\rangle|=|\det (\langle \Pi, \tilde{\Pi}_j\rangle)_{i,j=1}^d|=|\det \hat{O}(B_<,\tilde{B}_<)|,
\label{scalar}
\end{equation}
$B_<$ and $\tilde{B}_<$ are any ordered MORIs in the ${\cal S}_d$-equivalence class of $B=\{\Pi_i\}_{i=1}^d$ and $\tilde{B}=\{\tilde{\Pi}_i\}_{i=1}^d$ respectively.   From this it is evident that  (\ref{D_FS}) is nothing but the pull-back of the Fubini-Study
metric via the Pl\"ucker embedding (\ref{Plucker}). 

Unitary invariance of the metric (\ref{D_FS}) stems from the fact that
the overlap matrix associated with ${\cal U}({\cal A}_B)$ and ${\cal U}({\cal A}_{\tilde{B}})$ is given by
$\hat{O}(BU_<,\tilde{B}U_<)_{i,j}=\langle {\cal U}(\Pi_i), {\cal U}(\tilde{\Pi}_j)\rangle 
=\langle \Pi_i, \tilde{\Pi}_j\rangle=\hat{O}(B_<,\tilde{B}_<)_{i,j},\,(\forall i,j),$ where we have used unitary invariance of the Hilbert-Schmidt scalar product.

{\bf{iii)}} Finally, $\tilde{C}_{B}(U):=D_{FS}({\cal A}_B, {\cal U}({\cal A}_{B}))=0$ iff ${\cal A}_B={\cal U}({\cal A}_{B})$ iff $U$ is incoherent.
Moreover, if ${\cal W}$ is incoherent $\tilde{C}_{B}(WU):=D_{FS}({\cal A}_B, {\cal W}{\cal U}({\cal A}_{B}))=D_{FS}({\cal W}({\cal A}_B),
 {\cal W}{\cal U}({\cal A}_{B}))=D_{FS}({\cal A}_B, {\cal U}({\cal A}_{B}))=\tilde{C}_{B}(U),$ 
where we have used unitary invariance of $D_{FS}$. This shows that $\tilde{C}_{B}$ is a good CGP measure for unitaries \cite{CGP-0}.

Turning to $\varphi_B$ in (\ref{Cs}) we see that $\varphi_B(U)=0$ iff $|\det\hat{X}_{B_<}(U)|=1$ that from the above is equivalent to
$|\det \hat{O}(B_<, BU_<)|=1\Leftrightarrow B=BU$ namely  $U$ is incoherent. 
Since for  incoherent $W$'s one has $\hat{X}_{B_<}(WU)=Q_W \hat{X}_{B_<}$ (where $Q_W$ is a permutation matrix depending on $W$) one finds
$|\det\,\hat{X}_{B_<}(WU)|=|\det\,\hat{X}_{B_<}(U)|$ which shows  invariance under post-processing  by incoherent $W$'s.
$\hfill\Box$
\vskip 0.2truecm

In in view of Eq.~(\ref{ii}), and {\bf{Def. 5}}  one can  write $D^2({\cal A}_B,{\cal A}_{\tilde{B}})={2} (d-\|\hat{O}(B_<,\tilde{B}_<)\|_2^2)$
where $B_<$ and $\tilde{B}_<$ are any ordered MORIs in the ${\cal S}_d$-equivalence class of $B=\{\Pi_i\}_{i=1}^d$ and $\tilde{B}=\{\tilde{\Pi}_i\}_{i=1}^d.$  Eq.~(\ref{overlap-prop})] now shows that the maximum distance between MASAs is given by $\sqrt{2(d-1)}$ and it is achieved when the overlap matrix is given by the Van der Waerden's  matrix i.e., $\langle \Pi_i, \tilde{\Pi}_j\rangle=1/d, \,(\forall i,j)$. In this case the MORIs $B$ and $\tilde{B}$ correspond to {\em{mutually unbiased bases}} \cite{MUB} and the unitary connecting them, because of (\ref{C_B-dist}), has maximum CGP.

It is also worthwhile stressing that  Eq.~(\ref{scalar})  shows that the modulus of the determinant of the overlap matrix $\hat{O}(B_<,BU_<)=
\hat{X}_{B_<}(U)$  has a natural interpretation as {\em{fidelity}} between the input and output   MASAs ${\cal A}_B$ and ${\cal U}({\cal A}_B)$. 
%


 The second measure in Eq.~(\ref{Cs}) was introduced in \cite{CGP-1} [ii) in Prop. 9] here we see that it is rooted in the geometry of the Grassmannian seen as sub-variety of the
 projective space ${\mathbf{P}}(\bigwedge_{i=}^d {\cal H}_{HS})$. 
 If ${\cal H}={\cal H}_1\otimes {\cal H}_2,$ and $B_\alpha=\{\Pi^\alpha_i\}_{i=1}^{d_\alpha:={\mathrm{dim}}\,{\cal H}_\alpha}, (\alpha=1,2)$ 
 is a  MORI over ${\cal H}_{\alpha} ,\,(\alpha=1,2)$ one can define a product MORI by    $B=\{\Pi^1_i\otimes \Pi^2_j\,/\, i=1,\dots, d_1;  j=1,\dots, d_2\}$. 
From (\ref{additive}) (see also Ref.~\cite{CGP-1}) one finds
$\varphi_B(U_1\otimes U_2)=\varphi_{B_1}(U_1)+\varphi_{B_2}(U_2)$ 
where  $U_i\in U({\cal H}_i),\,(i=1,2)$ i.e., the measure $\varphi_B$  is {\em{additive}} \cite{additive}.

\section{Differential geometry of coherence power}
We now move to consider a differentiable metric structure. This is done in terms of the natural Rienmannian metric
over the Grassmannian $ds^2=D(P, P+dP)^2={\mathrm{Tr}}(dP^2)$. In view of the result (\ref{C_B-dist}) this metric will have the physical interpretation as the CGP of the unitary associated with an infinitesimal change of the MORI. For example if $B$ is the MORI associated to (a non-degenerate) Hamiltonian  $H$ a perturbation $H\mapsto H+\delta V=H^\prime$ will induce a change to an infinitesimally close one $B^\prime$. 
In view of Eq.~(\ref{C_B-dist}) the distance between the corresponding MASAs would then measure the CGP of the (infinitesimal) adiabatic intertwiner 
$\delta {\cal W}_{ad}$ between the eigenstate systems 
of $H$ and $H^\prime$  i.e., $C_B(\delta {\cal W}_{ad})=\frac{N_d}{2}\,ds^2$. 
\vskip 0.2truecm
{\bf{Proposition 6--}}
If $B=\{\Pi_i=|i\rangle\langle i|\}_{i=1}^d$ then 
\begin{equation}
ds^2={\mathrm{Tr}}(d{\cal D}_B)^2=4\sum_{i=1}^d \chi_i,\quad \chi_i:=\langle di|di\rangle-|\langle i|di\rangle|^2,
\label{diff-dist}
\end{equation}
Moreover $ds^2_{FS}=\frac{1}{2} ds^2$.

{\em{Proof.--}} 

Let us write differentials as $dX=\dot{X} dt$ then $(ds/dt)^2={\mathrm{Tr}}(\dot{{\cal D}_B})^2$.
One has $\dot{{\cal D}_B}=\sum_{i=1}^d (\dot{\Pi_i}\bullet {\Pi_i}+{\Pi_i}\bullet\dot{\Pi_i}),$
whence $(\dot{{\cal D}_B})^2=\sum_{i,j=1}^d\left( \dot{\Pi_j}(\dot{\Pi_i}\bullet {\Pi_i}+{\Pi_i}\bullet\dot{\Pi_i})\Pi_j
+{\Pi_j} (\dot{\Pi_i}\bullet {\Pi_i}+{\Pi_i}\bullet\dot{\Pi_i})\dot{\Pi_j}
  \right)$. Now, $ {\mathrm{Tr}}(\dot{{\cal D}_B})^2=2\sum_{i=1}^d {\mathrm{tr}}(\dot{\Pi_i})^2+
  \sum_{i,j=1}^d {\mathrm{tr}}(\Pi_j\dot{\Pi_i}){\mathrm{tr}}(\Pi_i\dot{\Pi_j}). $
  From orthornormality follows ${\mathrm{tr}}(\Pi_j\dot{\Pi_i})={\mathrm{tr}}(\Pi_j\dot{\Pi_i} \Pi_i)=0\,(\forall i,j),$
  therefore ${\mathrm{Tr}}(\dot{{\cal D}_B})^2=2\sum_{i=1}^d {\mathrm{tr}}(\dot{\Pi_i})^2$.
  By writing $\Pi_i=|i\rangle\langle i|$ and differentiating, a standard calculation \cite{DG-qpt} shows that 
  $\frac{1}{2}  \|\dot{\Pi_i}\|_2^2=\langle \frac{di}{dt}|\frac{di}{dt}\rangle -|\langle \frac{di}{dt}|i\rangle|^2=\chi_i/dt^2$ therefore
  $(ds/dt)^2=4 \sum_{i=1}^d( \langle \frac{di}{dt}|\frac{di}{dt}\rangle -|\langle \frac{di}{dt}|i\rangle|^2)$. Reabsorbing the $dt$ factors on the RHS
  one finds (\ref{diff-dist}).

To see that $ds^2_{FS}$ has the same expression as $ds^2$ we use the fact  $\det(\openone+\delta\hat{X}_B)= 1+{\mathrm{tr}}\,\delta\hat{X}+\cdots$.
Expanding $\tilde{\Pi}_j$ near $|\Pi_j$ in $(\hat{X}_B)_{ij}=\langle\Pi, \tilde{\Pi}_j\rangle$ one finds $(\delta\hat{X}_B)_{ij}=\langle\Pi_i, d{{\Pi}}_j\rangle
+\frac{1}{2} \langle\Pi_i, d^2{\Pi}_j\rangle$. Taking the trace one has $\sum_{i=1}^d(\delta\hat{X}_B)_{ii}=\frac{1}{2} \langle\Pi_i, d^2{\Pi}_i\rangle=
-\frac{1}{2}\sum_{i=1}^d \langle d\Pi_i, d{{\Pi}}_i\rangle$. Here we have used that $\Pi_i d\Pi_i \Pi_i=0$ and 
$\sum_{i=1}^d \langle\Pi_i, d^2{\Pi}_i\rangle=-\sum_{i=1}^d \|d\Pi_i\|_2^2$ (obtained by differentiating and adding the identities $\Pi_i^2=\Pi_i$.).  As in the above $\|d\Pi_i\|_2^2=2\chi_i$. Now one has that $ds^2_{FS}=\cos^{-1}(1-\sum_{i=1}^d\chi_i)$ and the claim is obtained by expanding the $\cos$. $\hfill\Box$  
\vskip 0.2truecm
The $\chi_i$'s in Eq.~(\ref{diff-dist}) are projective space metrics associated to the $|i\rangle$'s.
 When the latter are Hamiltonian eigenstates the $\chi_i$'s are known as fidelity susceptibilities.  The metric (\ref{diff-dist})
 is a sum of projective space ones. This reflects the fact that locally (see the numerator of Eq.~(\ref{M_d})) the set of MASAs
 is the full-flag manifold $U(d)/U(1)^d$ which is the set of ordered tuples $(\Pi_i)_{i=1}^d$. The latter can be 
 can be seen as a subvariety of ${\cal G}_1({\cal H})^d= {\mathbf{P}}({\cal H})^d$ by the obvious embedding.

Physically, the ground state susceptibility $\chi_0$  plays a key role in the differential geometric approach to quantum phase transitions (QPT) started in Ref. \cite{DG-qpt}. 
The idea is that when $\chi_0,$ which depends of the parameters defining the Hamiltonian, shows some singularity in the thermodynamical limit
or a super extensive (for local Hamiltonians) behaviour  for finite-size systems a QPT is occurring at that point in the parameter space \cite{DG-qpt}.

From this perspective  Eqs.~(\ref{C_B-dist}) and (\ref{diff-dist}) are intriguing as they comprise information about all eigenstates.  It is therefore tempting to wonder whether these geometric quantities, which are quantifying quantum coherence power at the same time, can be exploited to study phase transitions in which a radical change  is occurring at the level of {\em{whole Hamiltonian eigenstate system}} e.g., many-body localization
\cite{MBL-next}

\section{One Qubit}
In order to illustrate the general results proved in this paper we  consider explicitly the qubit case i.e., $d=2$. In this case   
\begin{equation}
X_2=\frac{U(2)}{U(1)\times U(1)}\cong \frac{SU(2)}{U(1)}\cong {\mathbf{CP}}^1\cong S^2 \Rightarrow{\cal M}_2\cong\frac{ S^2}{{\mathbf{Z}}_2},
\label{M_2}
\end{equation}
where we used Eq.~(\ref{M_d}) and   ${\cal S}_2\cong{\mathbf{Z}}_2$.
 This has a simple geometrical interpretation since  MORIs (and therefore MASAs)  in two-dimensions have the form
 $B=\{\Pi_\alpha\}_{\alpha=\pm}$ 
 where $\Pi_\alpha:= \frac{1}{2}(\openone +\alpha \,{\mathbf{n}}\cdot {\mathbf{\sigma}}), \,(\alpha=\pm)$ the $\sigma_\alpha$'s are the standard Pauli matrices and ${\mathbf{n}}=(n_x,n_y,n_z)\in{\cal S}^2$.  Thus it is clear that  ${\mathbf{n}}$ is identified with $-{\mathbf{n}}$ as they both correspond to the same MORI.
 This simple example also shows that the set of MASA may have non-trivial topology:
loops in ${\cal M}_2$ fall in two topologically distinct categories, the trivial (non-trivial)  which corresponds to $(\Pi_1,\Pi_2) \mapsto   (\Pi_1,\Pi_2)$ ($(\Pi_1,\Pi_2) \mapsto   (\Pi_2,\Pi_1)$)   i.e., $\pi_1({\cal M}_2)\cong {\mathbf{Z}}_2$.
 
  If  $\tilde{B}=\{\frac{1}{2}(\openone +\beta\, {\mathbf{\tilde{n}}}\cdot {\mathbf{\sigma}})\}_{\beta=\pm},$
one can easily check that the overlap matrix is given by 
$\hat{O}_{\alpha\beta}(B,\tilde{B})=\frac{1}{2}(1+\alpha\beta\,{\mathbf{n}}\cdot {\mathbf{\tilde{n}}}),\,(\alpha,\beta=\pm),$
whose spectrum is $\{1,\,{\mathbf{n}}\cdot {\mathbf{\tilde{n}}}\} $ and therefore ${\mathrm{det}}\,\hat{O}(B,\tilde{B})={\mathbf{n}}\cdot {\mathbf{\tilde{n}}},$  and $\|\hat{O}(B,\tilde{B})\|_2^2=1+({\mathbf{n}}\cdot {\mathbf{\tilde{n}}})^2$. Whence
\begin{equation}
D^2({\cal A}_B,{\cal A}_{\tilde{B}})=
{2} (d-\|\hat{O}(B,\tilde{B})\|_2^2)=
{2}(1-({\mathbf{n}}\cdot {\mathbf{\tilde{n}}})^2)=2\sin^2 \psi, 
\label{D_2}
\end{equation}
where $\psi:=\cos^{-1}({\mathbf{n}}\cdot {\mathbf{\tilde{n}}})$. 
From Eq.~(\ref{D_2}) we clearly see that the MASAs corresponding to 
${\mathbf{n}}$ and $-{\mathbf{n}}$ are identified thus confirming that globally ${\cal M}_2\cong S^2/{\mathbf{Z}}_2$ as given by Eq.~(\ref{M_2})\cite{locally}. On the other hand ${\mathbf{n}}\perp{\mathbf{\tilde{n}}}$  correspond to maximally far apart MASAs. 

Now we  consider the commutators $[\Pi_\alpha,\tilde{\Pi}_\beta]=\frac{i \alpha\beta}{2}  ({\mathbf{n}}\times  {\mathbf{\tilde{n}}})\cdot\sigma$.
From which $\sum_{\alpha,\beta=\pm} \|[\Pi_\alpha,\tilde{\Pi}_\beta]\|_2^2=2 \|{\mathbf{n}}\times  {\mathbf{\tilde{n}}}\|^2=2  \sin^2 \psi$.
Comparing this last Eq. with (\ref{D_2}) confirms {\bf{ii)}} of {\bf{Prop. 4}}. If ${\mathbf{n}}=(0,0,1)$ and $\tilde{B}=BU$ with $U= a|0\rangle\langle 0| +a^* |1\rangle\langle 1|- b^* |0\rangle\langle 1|+ b |1\rangle\langle 0|,\, (a=\cos(\theta/2),\, b=e^{i\varphi}\sin(\theta/2) )$. 
Then ${\mathbf{\tilde{n}}}=(\sin\theta\cos\varphi, \sin\theta\sin\varphi, \cos\theta)\Rightarrow\psi=\cos^{-1}({\mathbf{n}}\cdot {\mathbf{\tilde{n}}})=\theta$.  From {\bf{Prop.  4}} and (\ref{D_2}) one gets \cite{CGP-0}
 $$C_B(U)=\frac{N_d}{2} D^2({\cal A}_B, {\cal A}_{\tilde{B}})=\frac{1}{6} \sin^2\theta. $$
 Maximum CGP is attained by all $U$'s with $\theta=\pi/2$ irrespective of $\varphi$ as the corresponding ${\mathbf{\tilde{n}}}$' s are equidistant 
 from ${\mathbf{{n}}}=(0,0,1)$.

Furthermore,  from ${\mathrm{det}}\,\hat{O}(B,\tilde{B})={\mathbf{n}}\cdot {\mathbf{\tilde{n}}}=\cos \psi,$ it follows immediately from Eq.~(\ref{D_FS}) that 
$D_{FS}({\cal A}_B,{\cal A}_{\tilde{B}})=\cos^{-1}| \cos \psi|$  which is given by $\psi$ for $\psi\in[0,\pi/2]$ and by $\pi-\psi$ for $\psi\in[\pi/2,\pi]$. 
 
Finally, from $ds^2=2\sum_{\alpha=\pm}\|d\Pi_\alpha\|_2^2=2 \sum_{\alpha=\pm}\|\frac{\alpha}{2} d{\mathbf{n}}\cdot\sigma\|_2^2=
2\,\|d{\mathbf{n}}\|^2$ one sees that the CGP metric is proportional to the standard euclidean metric of $S^2$. 

\section{Conclusions}
In this paper we have unveiled  a deep connection between the notion of quantum coherence generating power of unitary operations 
(introduced in\cite{CGP-0,CGP-1} on purely physical grounds) and the geometry of the Grassmannian of subspaces of the algebra of linear operators.
Given a  maximal orthogonal resolution of the identity  $B$ in the Hilbert space ${\cal H }\cong {\mathbf{C}}^d$ of a quantum system one can consider the $d$-dimensional algebra ${\cal A}_B$ generated by $B$. 
This is a maximal abelian subalgebra (MASA) of the full operator algebra $L({\cal H})$ which is closed under hermitean conjugation.
The set of all MASAs is a topologically non-trivial subset of the Grassmannian  of $d$-dimensional subspaces of the Hilbert-Schmidt space $L({\cal H})\cong {\mathbf{C}}^{d^2}$.
Given a unitary map $\cal U$ we have shown  that its coherence  generating power  with respect $B$ is proportional to the Grassmannian distance, as well as the lack of commutativity, between the MASAs  ${\cal A}_B$ and ${\cal U}({\cal A}_B)$.  By embedding the set of MASAs  into the projectivation of the $d$-th exterior power of Hilbert-Schmidt space one can pull-back the standard Fubini-Study metric and obtain novel coherence power measures endowed  by a natural geometrical interpretation. 

\begin{acknowledgments}
This work was partially supported by the ARO MURI grant W911NF-11-1-0268. We thank G. Styliaris for useful discussion and for insisting on the "geometrical approach".
\end{acknowledgments}


\begin{thebibliography}{100}
\bibitem{refine} $B^\prime=\{P_j\}_{j=1}^{m^\prime}$ is refinement of $B=\{Q_j\}_{j=1}^{m}$ iff $m^\prime\ge m$
and it exists a partition of ${1,\ldots,m^\prime}$ in $m$ disjoint subset ${\cal S}_\alpha\,(\alpha=1,\ldots,m)$ such that $ P_\alpha= \sum_{j\in{\cal S}_\alpha} Q_j\,(\forall\alpha)$.
\bibitem{uni-inv} In fact the projection associated to ${\cal U}({\cal A}_B)$ is given by ${\cal U} {\cal D}_B {\cal U}^\dagger$
whence $D({\cal U}({\cal A}_B), {\cal U}({\cal A}_{\tilde{B}}))=\|{\cal U} {\cal D}_B {\cal U}^\dagger-{\cal U} {\cal D}_{\tilde{B}} {\cal U}^\dagger\|_{HS}
= \|{\cal U} ( {\cal D}_B - {\cal D}_{\tilde{B}}) {\cal U}^\dagger\|_{HS}=\| {\cal D}_B - {\cal D}_{\tilde{B}}\|_{HS}= 
D({\cal A}_B, {\cal A}_{\tilde{B}})$. Where we have used the unitary invariance of the Hilbert-Schmidt norm.
\bibitem{flag} An orthogonal  full-flag is a sequence of
 subspace  $V_1\subset V_2\subset\cdots\subset V_d={\cal H}$ such that 
${\mathrm{dim}}(V_k)=k,\,V_{k+1}=(V_k^\perp\cap V_{k+1})\oplus V_k,\,(k=1,\ldots,d-1)$ or,  equivalently, a sequence of $1$-dimensional orthogonal subspaces $(W_k)_{k=1}^d$ such that $V_k=\oplus_{j=1}^k W_j$. In our case  
$W_j={\mathrm{Im}\,} \Pi_j,\,(j=1,\dots,d)$.
\bibitem{commA-B} Notice that if ${\cal A}$ and ${\cal B}$ are MASAs  then ${\cal A}={\cal B}$ iff $[{\cal A}, {\cal B}]=0$.
Indeed if the latter equality is true one has that ${\cal B}\subset {\cal A}^\prime$($=$ commutant of $\cal A$). But ${\cal A}={\cal A}^\prime$
from maximality. Therefore ${\cal B}\subset{\cal A}$. The converse inclusion is obtained in the same way and therefore ${\cal A}={\cal B}$.
\bibitem{ext} If $W$ is a $n$-dimensional  linear space (with $n\ge d>0$)we denote  by $\bigwedge_{k=1}^d W$ the total anti-symmetric part of $W^{\otimes\,d}$. By ${\mathbf{P}}(W)$ we denote the projective space associated with $W$.
\bibitem{additive} Additivity is a direct consequence of the property $X_B(U_1\otimes U_2)= X_{B_1}(U_1)\otimes X_{B_2}(U_2)$ and 
$\det (A\otimes B)= \det(A)^{d_2} \det(B)^{d_1}$.
\end{thebibliography}
\end{document}